\documentclass[final]{svjour2}
\usepackage{graphicx}
\usepackage{rotating}
\usepackage{amssymb}
\usepackage{mathptmx}
\usepackage{color}
\usepackage[numbers]{natbib}
\makeatletter
\journalname{Journal of Low Temperature Physics}

\bibpunct{}{}{,}{s}{}{,}

\begin{document}

\newcommand{\hdblarrow}{H\makebox[0.9ex][l]{$\downdownarrows$}-}
\title{A Titanium Nitride Absorber for Controlling Optical Crosstalk in
  Horn-Coupled Aluminum LEKID Arrays for Millimeter Wavelengths}

\author{H.~McCarrick$^{1,a}$ 
  \and D.~Flanigan$^1$
  \and G.~Jones$^1$
  \and B.~R.~Johnson$^1$
  \and P.~A.~R.~Ade$^2$
  \and K.~Bradford$^2$
  \and S.~Bryan$^2$
  \and R.~Cantor$^6$
  \and G.~Che$^2$
  \and P.~Day$^3$
  \and S.~Doyle$^5$
  \and H.~Leduc$^3$
  \and M.~Limon$^1$
  \and P.~Mauskopf$^2$
  \and A.~Miller$^1$
  \and T.~Mroczkowski$^7$
  \and C.~Tucker$^5$
  \and J.~Zmuidzinas$^{3,4}$
}


\institute{
1) Department of Physics, Columbia University, New York, NY 10025, USA \\
2) Department of Physics, Arizona State University, Tempe, AZ 85287, USA \\
3) Jet Propulsion Laboratory, Pasadena, CA 91109, USA \\
4) Caltech, Pasadena, CA 91109, USA \\
5) School of Physics and Astronomy, Cardiff University, Cardiff, Wales CF24 3AA, UK \\
6) STAR Cryoelectronics, Santa Fe, NM 87508, USA \\
7) Naval Research Laboratory, Washington, DC 20375, USA \\
a) email: hlm2124@columbia.edu \\
}

\maketitle


\begin{abstract}

We discuss the design and measured performance of a titanium nitride (TiN) mesh absorber we are developing for controlling optical crosstalk in horn-coupled lumped-element kinetic inductance detector arrays for millimeter-wavelengths.
This absorber was added to the fused silica anti-reflection coating attached to previously-characterized, 20-element prototype arrays of LEKIDs fabricated from thin-film aluminum on silicon substrates.
To test the TiN crosstalk absorber, we compared the measured response and noise properties of LEKID arrays with and without the TiN mesh.
For this test, the LEKIDs were illuminated with an adjustable, incoherent electronic millimeter-wave source.
Our measurements show that the optical crosstalk in the LEKID array with the TiN absorber is reduced by~66\% on average, so the approach is effective and a viable candidate for future kilo-pixel arrays.  
%
%


\end{abstract}


\section{Introduction}

A lumped-element kinetic inductance detector (LEKID) is a superconducting, photon-sensing resonator consisting of a capacitor and an inductor.
The inductance has both geometric and kinetic components, the latter arising in alternating currents only and produced by stored energy in the Cooper pairs.
When a photon with energy above the gap energy of the detector material is absorbed, the resonance frequency $f$ and quality factor $Q$ of the resonator shift\cite{day, zmu}.
The detector is coupled to a transmission line, allowing these perturbations to be measured with probe tones.

\begin{figure}
\centering
\includegraphics[width=\textwidth]{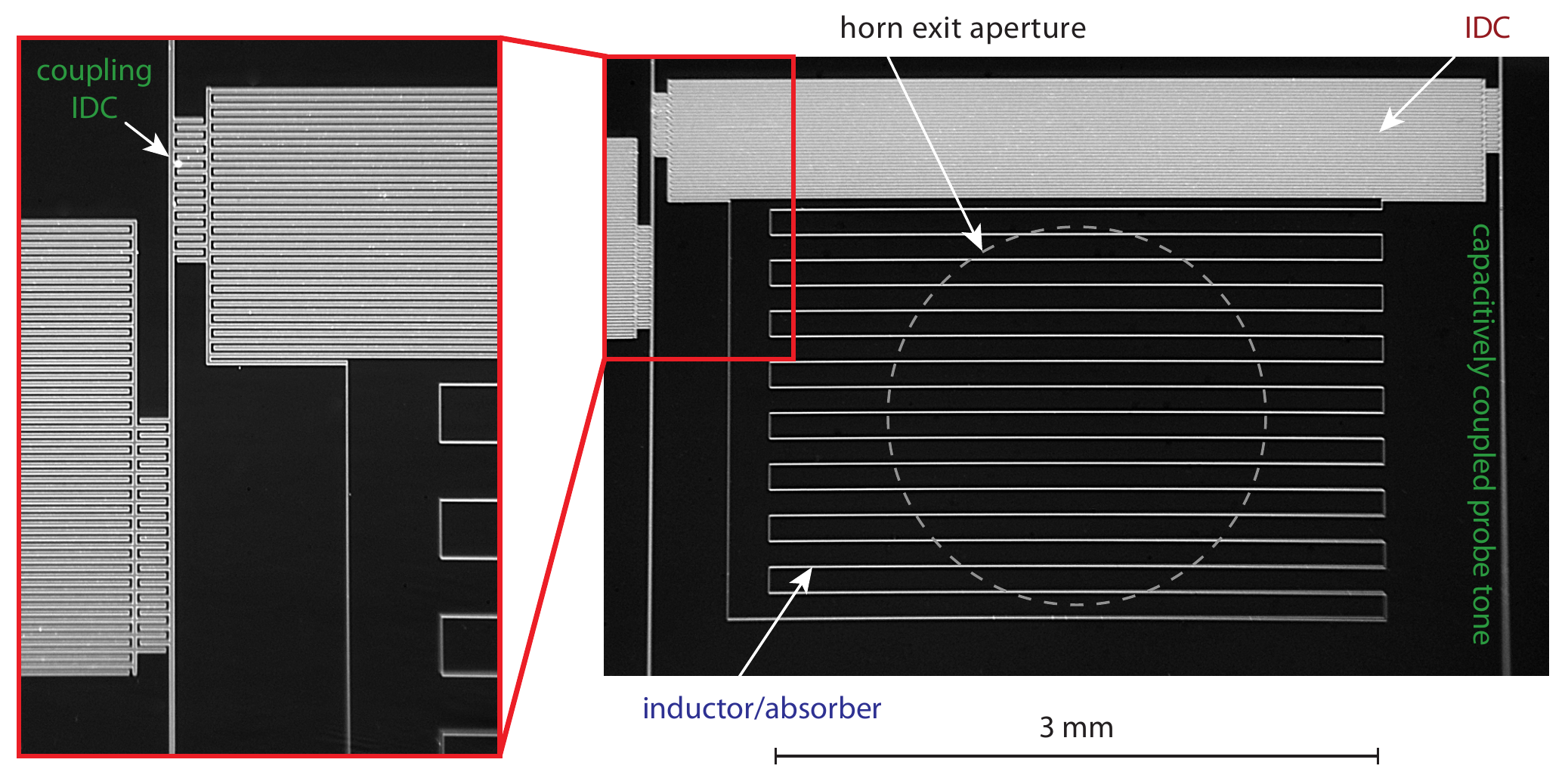}
\caption{
A photograph of a LEKID used in this study. The interdigitated capacitor (IDC) and meandered inductor set the resonance frequency for each detector. Incident photons are absorbed in the inductor and break Cooper pairs changing the kinetic inductance. This shifts the resonance frequency of the detector. The inset shows the IDC that couples the resonator to the transmission line, allowing the frequency perturbation to be readout. (Color figure online.)
}
\label{detector}
\end{figure}

\begin{figure}
\centering
\includegraphics[width=\textwidth]{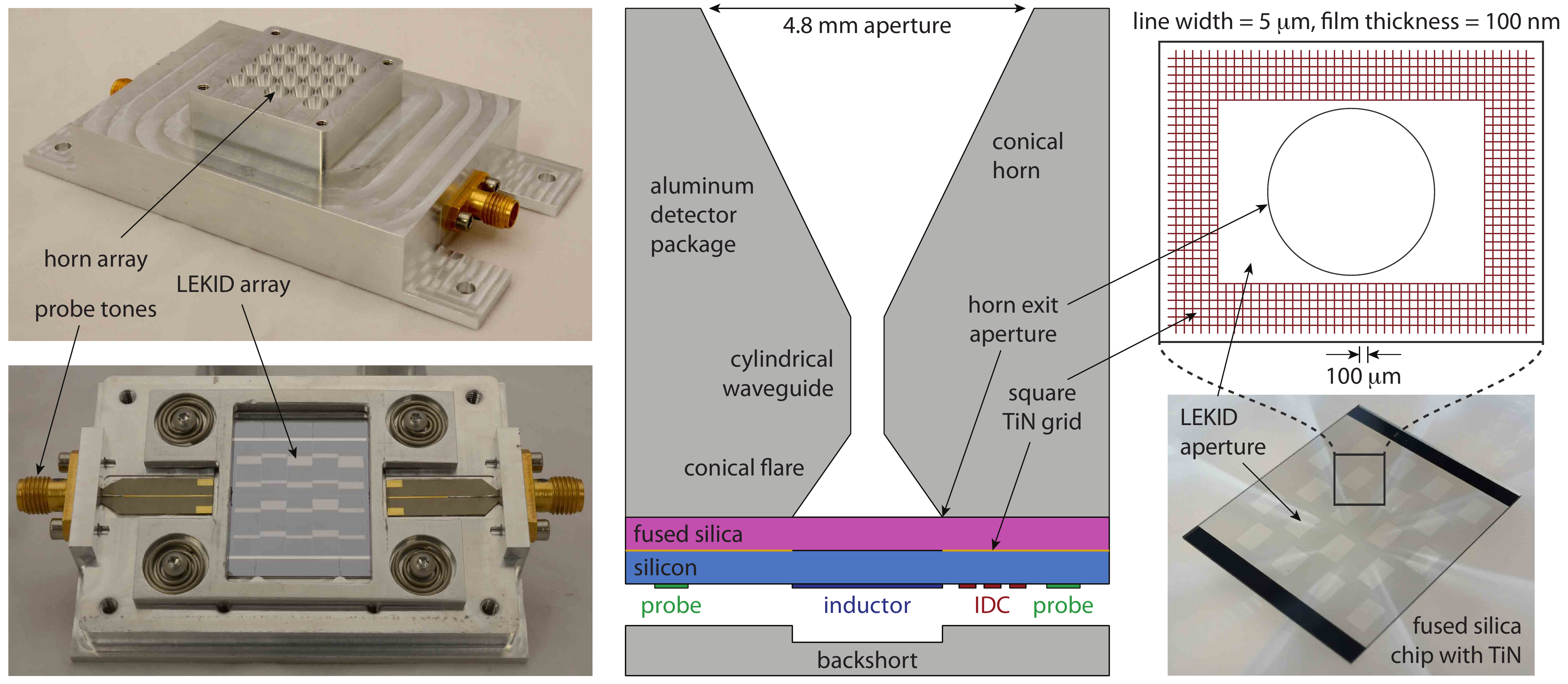}
\caption{
\textit{Left Top:} Photograph of the conical horns machined into the aluminum detector package. 
\textit{Left Bottom:} Photograph of the LEKID array mounted in the detector package with the backshort plate removed. 
\textit{Center:} Cross-sectional view of one array element.
\textit{Right Top}: Schematic of the TiN absorber designed to mitigate optical crosstalk between detectors. The 100~nm thick TiN absorber is fabricated on a fused silica chip, which is nominally used for matching the impedance of the horn to the LEKID array. Rectangular apertures in the mesh the same size as the LEKID inductor/absorber allow photons to propagate from the horn to the detector.
\textit{Right Bottom:} Photograph of the TiN absorber fabricated at NASA/JPL. (Color figure online.)
}
\label{detectors}
\end{figure}

In this paper, we present the measured performance of a titanium nitride (TiN) mesh designed to control optical crosstalk in arrays of horn-coupled LEKIDs.
The 20-element prototype LEKID arrays used in this study~\cite{mccarrick14} are sensitive to a 40~GHz spectral band centered on 150~GHz.
These LEKIDs are being developed for cosmic microwave background (CMB) studies at millimeter wavelengths.
The LEKIDs and the TiN mesh are shown in Figs.~\ref{detector}~and~\ref{detectors}.
The new TiN absorber introduced in this work enhances the performance of these LEKIDs by absorbing photons propagating laterally in the dielectrics inside the detector package.
Therefore, the TiN mesh absorbs the photons that produce optical crosstalk.


\section{Experiment Details}


Each LEKID is coupled to a conical horn, as shown in Fig.~\ref{detectors}. 
A low-pass metal mesh filter~\cite{ade06} mounted before the aperture of the conical horn defines the 170~GHz, high-frequency edge of the spectral band.
The conical horn flare tapers to a cylindrical, single-moded waveguide, and this waveguide acts as a high-pass filter, which defines the 130~GHz, low-frequency edge of the spectral band.
The cylindrical waveguide then expands into a second conical flare that is used to help match the wave impedance between the waveguide and the LEKID. 
Further, a 300~$\mu$m thick fused silica layer was inserted between the two elements to serve as an anti-reflection (AR) coating.
The back-illuminated LEKID arrays were fabricated from a 20~nm thick aluminum film patterned on a 300~$\mu$m thick high-resistivity, float-zone silicon wafer.
The detectors consist of a meandered inductor and interdigitated capacitor (IDC).
Each inductor in the array is identical.
Therefore, the unique resonance frequency of each detector is set by the capacitance of the IDC, which varies from device to device.
The detectors are operated at approximately 120~mK using a pulse-tube cooler and a two-stage adiabatic demagnetization refrigerator (ADR).
We use an FPGA-based digital readout system built from a ROACH signal processing board, which provides multiple, parallel homodyne readout chains.

Simulations show that photons can propagate laterally in the two dielectrics, and these photons could produce unwanted optical crosstalk.
Therefore, we implemented a 100~nm thick TiN mesh deposited on the detector side of the AR coating.
The TiN mesh is designed to have a wave impedance $Z = 194~\Omega$ matching that of the fused silica substrate.
To get this impedance, we patterned a mesh in the TiN film.
The line width of the TiN in the mesh is 5~$\mu$m, and the cell size is 100~$\mu$m by 100~$\mu$m.
The mesh is approximately the same size as the LEKID array (22~mm by 26~mm), and rectangular apertures 2~mm by 3~mm are patterned in the mesh in front of each inductor, so the signal photons can pass from the horn to the absorber.
Millimeter-wave photons traveling laterally along the dielectrics should be absorbed by this TiN mesh because the measured $T_c$ of the film is 1~K, so the gap frequency is below the passband of the detector module.

The optical response of the detectors was measured using an electronic photon source that produces broadband, incoherent millimeter-wave signals.
In this source, thermal noise from a 50~$\Omega$ termination resistor is first amplified in the 12~GHz range and then used to drive a 12$\times$ active frequency multiplier to produce 140 to 160~GHz radiation in a WR6 rectangular waveguide.
The millimeter-wave signal can be amplitude modulated using a PIN switch at the input of the multiplier.
Source signals are coupled into the cryostat through a WR6 waveguide window.
Inside the cryostat, a horn and a reflective collimator are used to illuminate the detectors.
The source is described in more detail in the literature~\cite{jones15, flanigan15}.
The power incident on the detectors was swept from approximately 0.1~to~20~pW to measure the responsivity of the devices.
The source power was varied using inline WR6 attenuators mounted outside the cryostat.

\begin{figure}
\centering
\includegraphics[width=\textwidth]{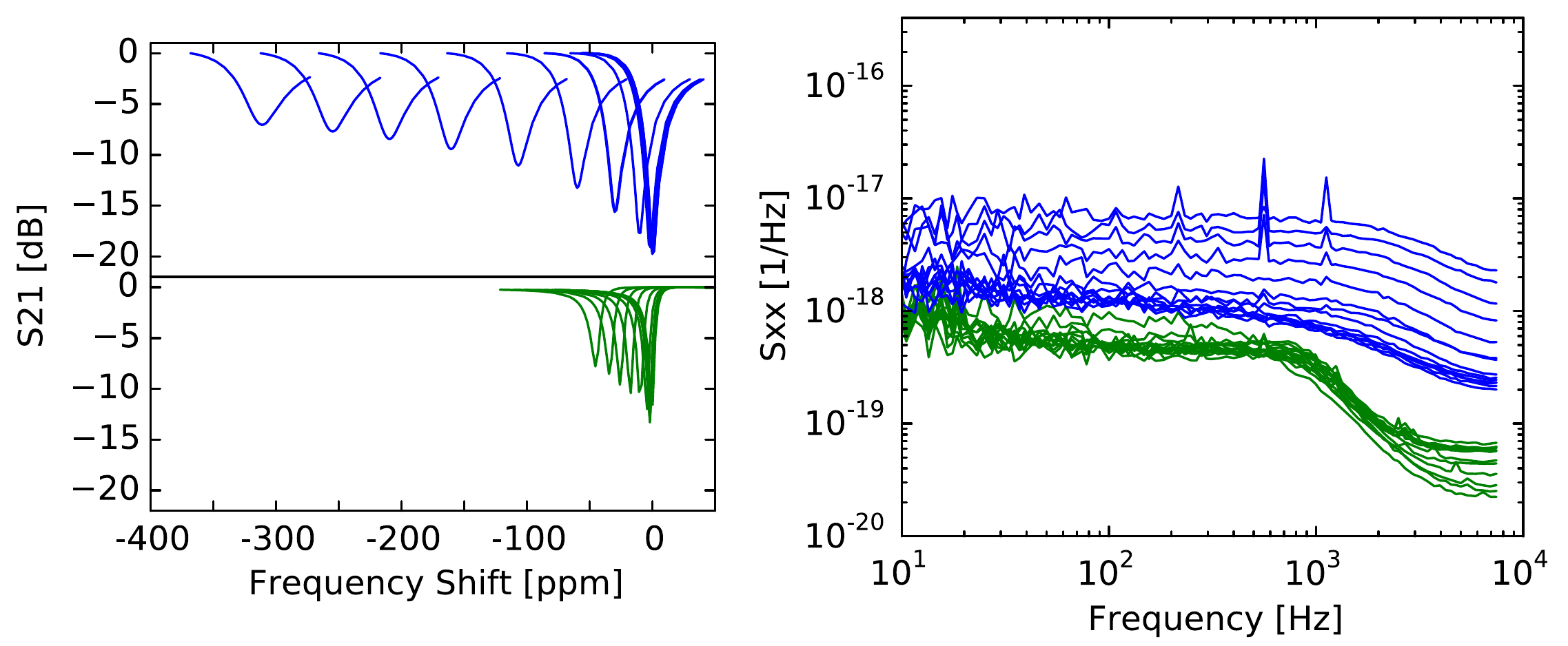}
\caption{
\textit{Left:} Measured S21 with changing optical load for the illuminated detector {\it blue} and a representative dark detector \textit{green} before the TiN mesh was added. The optical load was swept from approximately 0.1~to~20~pW. The ambient background loading in the test setup dominates at the low end of this range.
\textit{Right:} The noise spectrum of the illuminated detector \textit{blue} and a representative dark detector \textit{green} measured with different optical loads and constant probe-tone power. The optical power again varies from from approximately 0.1~to~20~pW. For the illuminated detector the device noise level (below 500 Hz) rises with optical power as the quality factor decreases. For the dark detector the device noise stays relatively constant with an increasing optical load, as expected. The roll off at 1~kHz is due to the resonator ring-down time. The noise level above the roll off is set by the amplifier noise. For clarity, the 60~Hz line and harmonics were removed. (Color figure online.)
}
\label{responsivity}
\end{figure}

\section{Results}


For this experiment, two different arrays of the same design, processed on the same wafer, were measured during two separate cryostat cycles.
One assembly included the TiN absorber and one did not. 
All of the horn apertures were covered with aluminum tape except for one allowing just one directly illuminated detector in the study.
We will henceforth refer to the taped-over horns as ``dark.''

\begin{figure}
\centering
\includegraphics[width=\textwidth]{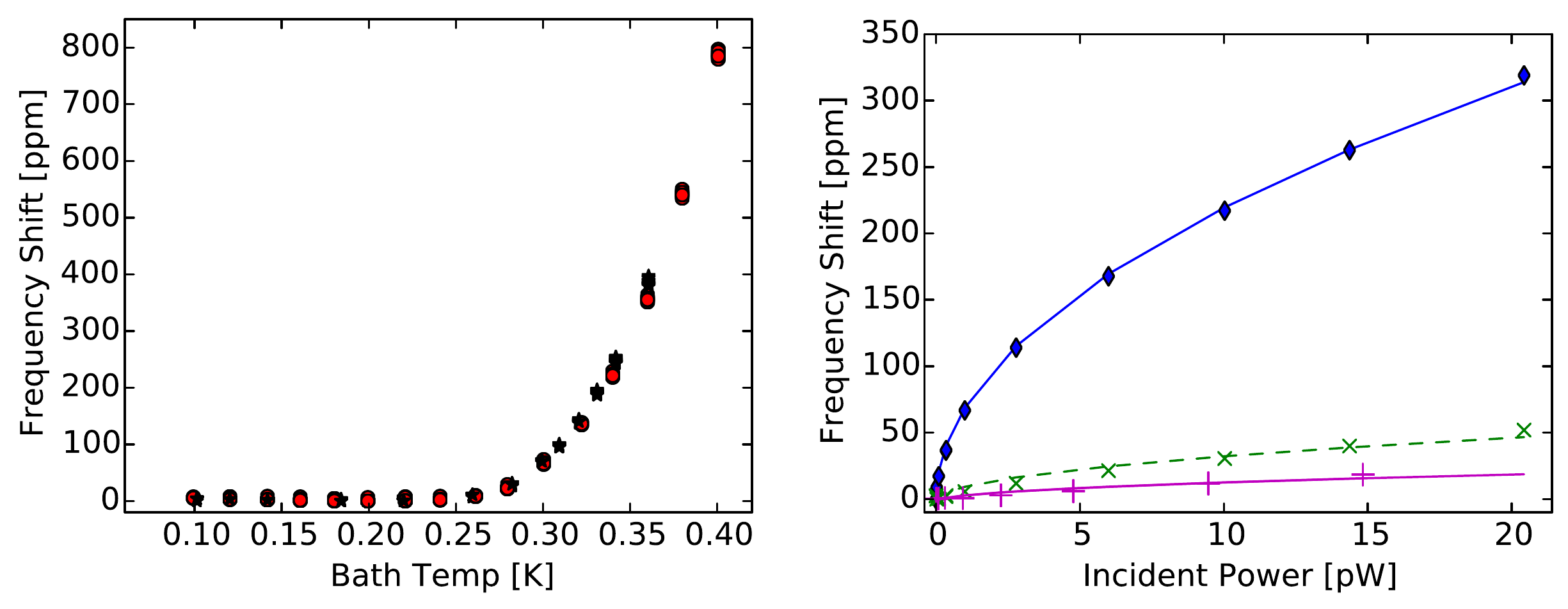}
\caption{
\textit{Left:} The fractional frequency change versus bath temperature for the detectors in the two arrays we used for this study.  The bath temperature was stepped from 100~mK to 400~mK to check for consistency in device responsivity between the arrays. The detectors without the TiN mesh are plotted as the red circles, and the detectors with the TiN mesh are plotted as the black stars.  
\textit{Right:} The fractional frequency change as a function of optical power for the illuminated detector (blue), a representative dark detector without the TiN mesh (green), and a representative dark detector with the TiN mesh (magenta). (Color figure online.) 
}
\label{bath}
\end{figure}

The measured response of both the illuminated 179~MHz resonator and a representative dark resonator in the TiN-free assembly is shown in Fig.~\ref{responsivity}.
For the assembly with the TiN absorber, the 179~MHz resonator in the array was not working.
Therefore, we instead illuminated the 102~MHz resonator. 
We checked that the responsivity to quasiparticles was consistent across all detectors in both arrays by measuring their response to changes in bath temperature.
As shown in Fig.~\ref{bath}, the response of the detectors across both arrays is nearly identical, indicating the responsivity of each detector in the test is the same and the two arrays can be meaningfully compared. 

\begin{figure}
\centering
\includegraphics[width=\textwidth]{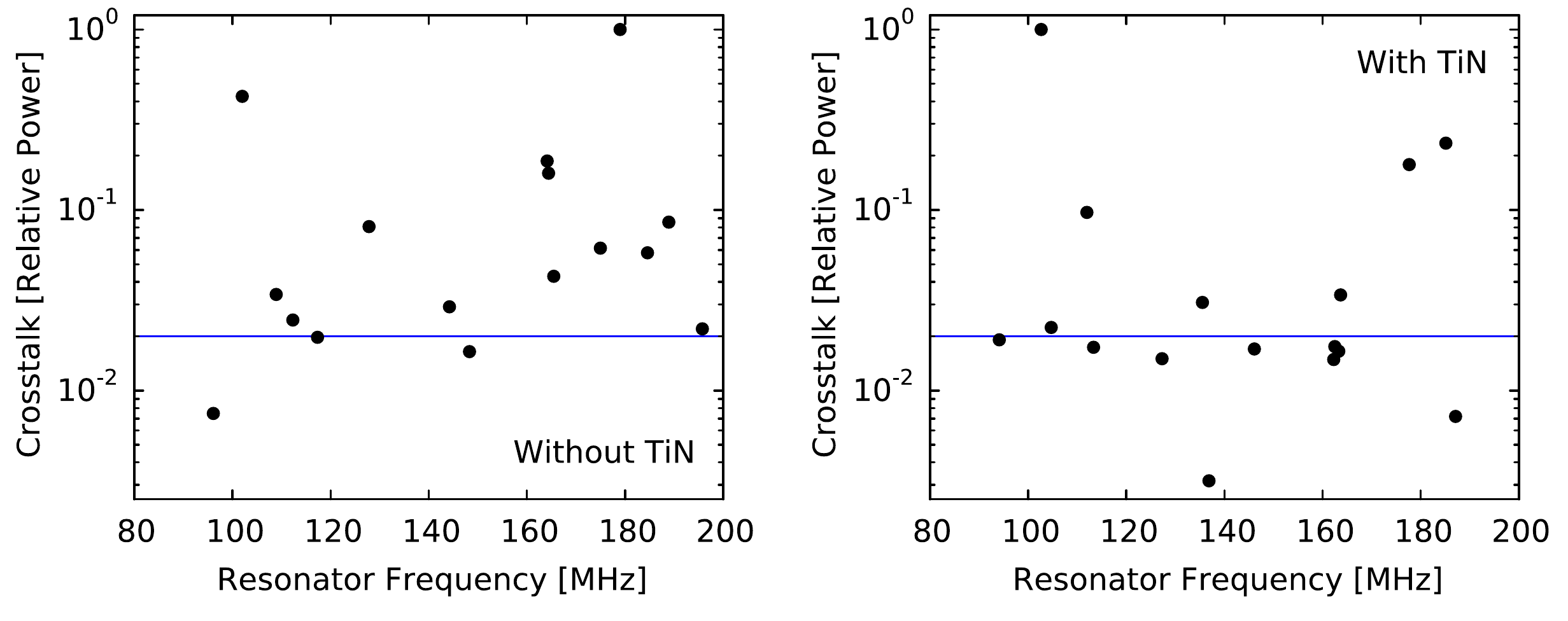}
\caption{
\textit{Left:} The power absorbed by each LEKID normalized to the power absorbed by the illuminated LEKID.  The module configuration used to produce the result in this panel did not include the TiN absorber. Here the illuminated LEKID absorbed approximately 0.3~pW of power.
\textit{Right:} The same result as that plotted in the left panel, though to produce the result in the right panel, the TiN absorber was used. The TiN absorber reduces the optical crosstalk on average by 66\%.  The detectors adjacent to the illuminated detector still show appreciable crosstalk response, which indicates our prototype detector module design needs to be optimized (see discussion in Section~\ref{discussion}).
The blue reference line in both panels is plotted at the 2\% level. (Color figure online.)
}
\label{tin}
\end{figure}

Because the optically-produced quasiparticles dominate, the resonator frequency response $\delta f/f$ should be well described by a function proportional to $(1 + P/P_*)^{1/2} - 1$, where $P$ is the incident power and $P_*$ is a film-dependent constant \cite{mccarrick14,mckenney2012}. 
We originally fit this model to the illuminated detector response, and used the best-fit model parameters to compute the relative power absorbed by the dark detectors (optical crosstalk).
We found this method underestimates the amount of power absorbed in the dark detectors because the fit in the lower-power regime ($\ll 1$~pW of incident power) is poor, likely due to systematic errors produced by the ambient background signal dominating the comparatively small input test signal from our source.
Therefore, instead we measured the frequency shifts with 0.3~pW of loading.
This loading level should be in the linear regime of the detector response function, meaning the measured frequency shift in parts-per-million [ppm] should be directly proportional to power.
We then calculate the optical crosstalk by normalizing the measured frequency shifts to the frequency shift of the illuminated detector.
This result serves as an upper bound on the optical crosstalk because any non-linear LEKID response would compress the signal and spuriously decrease the normalization factor. 
We find that the optical crosstalk with the TiN absorber is lower than the optical crosstalk without the TiN absorber by 66\% on average as seen in Fig~\ref{tin}.
The detectors adjacent to the illuminated detector still show appreciable optical crosstalk that is reduced but not eliminated by the TiN absorber (see discussion in Section~\ref{discussion}).
And the response of the dark detectors not adjacent to the illuminated detector in the module with the TiN absorber is similar to the measured response of a third background/control configuration where all the horns are covered with aluminum tape; the TiN absorber reduces the optical crosstalk to a level that cannot be distinguished from this small systematic error background at the $\sim$1\% level.


\section{Discussion}
\label{discussion}

This study was our first systematic investigation into the level of optical crosstalk in our prototype horn-coupled LEKID modules. 
Our data reveals that the TiN absorber works well, and it reduces the apparent optical crosstalk to approximately 2\% or below for the detectors not adjacent to the illuminated detector in the array.
This level is comparable to the magnitude of systematic errors produced by the experimental setup, so further investigation is required to understand the optical crosstalk level more precisely; the measured $\sim$2\% crosstalk level is likely an upper limit.
Electromagnetic simulations show that approximately 5\% of the power from a horn should propagate to the surrounding devices without the TiN mesh, and approximately 1\% after adding the TiN mesh. 
The detectors adjacent to the illuminated detector have a somewhat higher than expected level of optical crosstalk that is reduced but not eliminated by the TiN absorber.
This crosstalk signal is likely produced by the detector module design and not a flaw in the LEKIDs themselves.
The prototype detector module presented here is our first design, and it requires further optimization.
To decrease the crosstalk further we will (i) add an RF choke around the exit aperture of the horn and/or the backshort cavity and (ii) decrease the gap between the LEKID array and the backshort plate to make it more difficult for light to propagate laterally in this space.
Prospects for decreasing ambient pickup further include improving the package seal to make it more light tight and using additional filters on the coaxial line.

%






\begin{acknowledgements}

This research is supported in part by a grant from the Research Initiatives for Science and Engineering (RISE) program at Columbia University to Bradley Johnson, and by internal Columbia University funding to Amber Miller. Heather McCarrick is supported by a NASA Earth and Space Science Fellowship (NESSF). We thank the Xilinx University Program for their donation of FPGA hardware and software tools used in the readout system.

\end{acknowledgements}



\end{document}